\documentclass[12pt,a4paper,final]{iopart}
\usepackage{iopams}  

\usepackage[utf8]{inputenc} % allow utf-8 input
\usepackage[T1]{fontenc}    % use 8-bit T1 fonts
\usepackage{url}            % simple URL typesetting
\usepackage{booktabs}       % professional-quality tables
\usepackage{amsfonts}       % blackboard math symbols
\usepackage{nicefrac}       % compact symbols for 1/2, etc.
\usepackage{microtype}      % microtypography
\usepackage{lipsum}		% Can be removed after putting your text content

\usepackage{color}
\usepackage{graphicx}
\newcommand{\cor}[1]{\textcolor{black}{#1}}
\DeclareUnicodeCharacter{2009}{\,} 
\begin{document}

\title[Imaging three-dimensional magnetic systems with X-rays]{{Imaging} three-dimensional magnetic systems with X-rays}

\author{C. Donnelly$^{1}$}
\address{$^1$Cavendish Laboratory, University of Cambridge, JJ Thomson Ave, Cambridge CB3 0HT, UK}
\ead{cd691@cam.ac.uk}

\author{V. Scagnoli$^{2,3}$}
\address{$^2$Laboratory for Mesoscopic Systems, Department of Materials, ETH Zurich, 8093 Zurich, Switzerland}
\address{$^3$Laboratory for Multiscale Materials Experiments, Paul Scherrer Institute, 5232 Villigen PSI, Switzerland}
\ead{valerio.scagnoli@psi.ch}

\begin{abstract}
Recent progress in nanofabrication and additive manufacturing have facilitated the building of nanometer-scale three-dimensional structures, 
that promise to lead to an emergence of new functionalities within a number of fields, compared to state-of-the-art two dimensional systems. In magnetism, the move to three-dimensional systems offers the possibility for novel magnetic properties not available in planar systems, as well as enhanced performance, both of which are key for the development of {new} technological applications.
In this review paper we will focus our attention on  three-dimensional magnetic systems and how their magnetic configuration can be retrieved using X-ray magnetic nanotomography.
We will start with an introduction to magnetic materials, and their relevance to our everyday life, along with  the growing impact that they will have in the incoming years in, for example, reducing energy consumption. We will then briefly introduce common methods used to study magnetic materials, such as electron holography, neutron and X-ray imaging. In particular, we will focus on X-ray magnetic circular dichroism and how it can be used to image magnetic moment configurations.
As a next step we will introduce  tomography  for three-dimensional imaging, and how it can be adapted to study magnetic materials. Particular attention will be given to explaining the reconstruction algorithms that can be used to retrieve the magnetic moment configuration from the experimental data, as these represent one of the main challenges so far, as well as the different experimental geometries that are available. Recent experimental results will be used as specific examples to guide the reader through each step in order to make sure that the paper will be accessible for those interested in the topic that do not have a specialized background on magnetic imaging.
Finally, we will describe the future prospects of such studies, identifying the current challenges facing the field, and how these can be tackled. In particular we will highlight the exciting possibilities offered by the next generation of synchrotron sources which will deliver diffraction limited beams, as well as with the extension of well-established methodologies  currently implemented for the study of two-dimensional magnetic materials to achieve higher dimensional investigations.
\end{abstract}

%Uncomment for PACS numbers title message
% Keywords required only for MST, PB, PMB, PM, JOA, JOB? 
\vspace{2pc}

% Uncomment for Submitted to journal title message
\submitto{\JPCM}
% Comment out if separate title page not required
%\maketitle

\section{Introduction}
Magnetic materials have played a pivotal role in the industrialization of our society. Initially with dynamos, and subsequently alternators, which all made use of permanent magnets, new capabilities for transforming electrical power into kinetic energy, and vice-versa, led to \textit{Electrification},  hailed as ``the greatest engineering achievement of the 20th Century'' \cite{twenty_century}. 
The importance of such devices, whose performance has been improved tremendously in the last decades, has not faded with time. On the contrary, today permanent magnet-based devices form the basis of renewable energy production, forming a main component of, for example, wind turbines, \cor{and} represent a key component of the automotive industry, where complex stacks of magnetic and non-magnetic layers are also used as sensors \cite{TREUTLER20012}.

Aside from energy production and harvesting, one of the most well-known applications of magnetism in our day-to-day lives is magnetic recording media, which until recently was the main way to store information in our personal computers. Although personal storage on mobile devices has shifted towards solid-state drives in recent years, magnetic recording media is still used extensively in international data storage centres, such as those serving \textit{the Cloud}.  
As the amount of information to be stored or transmitted increases, so does the pressure to develop and provide high capacity storage at low economic costs, whilst growing awareness of the {high}  energy consumption and carbon footprint of data storage centres \cite{mills13} demands  more energy-efficient, sustainable storage options. Currently there is significant research  into developing new energy-saving materials for information technology \cite{non_volatile_memory},  such as multiferroics  \cite{gajek07,bibes08,Pyatakov_2012,spaldin19}, skyrmions \cite{fert13}, and magnetic \cite{kent15} and antiferromagnetic \cite{jungwirth16,jungwirth18} spintronic devices, all realised by the development of complex magnetic materials.

%As well as the development of new materials, in recent years it has been shown that enhanced, or indeed new, physical properties and functionality can also be  obtained through an increase in the dimensionality of a system.  
An alternative route to enhanced, or indeed new, physical properties and functionality can also be  obtained through an increase in the dimensionality of a system. Indeed, in recent years the introduction of three-dimensionality has led to advances in a number of fields, including significant increases in  the energy density of  solar cell technologies with the macroscopic three-dimensional arrangement of solar cells \cite{myers10,bernardi12},  increased control over both photonic \cite{gansel09,Yang10,blanco00} and mechanical \cite{bueckmann12} properties in three-dimensional metamaterials, and new medical applications such as drug delivery with fabricated three-dimensional magnetic microbots \cite{tottori12,nelson10}.  

When it comes to magnetic materials,  three-dimensionality is no less promising. Two recent review papers~\cite{fernandezpacheco17,streubel16} highlighted the wide-ranging potential of curved and three-dimensional magnetic systems, from opportunities for ultra-high density data storage \cite{parkin08}, to new intrinsic geometry-induced magnetic properties \cite{sheka15,gadidei14,Pylypovskyi15} and extraordinary magnetisation dynamics \cite{hertel16}. Now, with recent advances in modelling and synthesis, the design and fabrication of three-dimensional magnetic nanosystems has become possible \cite{deteresa16,fernandezpacheco13,fernandezpacheco09,Sanz-Hernandez18_ACSNano,May19,vavassori16,keller18}, opening the door to a wide variety of new physics, and future applications.

As well as new fabrication and simulation expertise, robust characterisation techniques well-suited to these complex devices and geometries are essential to develop a fundamental understanding of the behaviour of these systems.
Recent advances in magnetic tomography have opened the door to the detailed investigation of three-dimensional magnetisation configurations \cite{streubel15,donnelly17,tanigaki15,manke10,kardjilov08,Wolf2019,wolf2015,hilger2018,HierroRodriguez19}, which is key to the development of current and future magnetic devices. In this review we will discuss the promise of three-dimensional magnetic systems, and recent developments for their characterisation, with a particular focus on the promising area of X-ray magnetic tomography.

This review is organized as follows: in Section~\ref{sec:3Dmag} we provide an introduction to three-dimensional magnetic systems. Section~\ref{sec:Char3DSt} gives a general introduction on the characterization of three-dimensional magnetic nanostructures, with an emphasis on X-ray techniques. Section~\ref{sec:xrayIm} describes in more detail X-ray magnetic imaging in two dimensions, illustrating the advantages and disadvantages of  different X-ray imaging techniques. In Section~\ref{sec:3DIm} we explore three-dimensional imaging  of magnetic materials with X-rays, describing in detail both the state-of-the-art, and the needs and requirements for the experiments. Finally, in Section~\ref{sec:FutPr} we discuss the prospective technical improvements for X-ray three-dimensional imaging and interesting scientific cases that could be investigated by this technique. 

\section{\label{sec:3Dmag}Three-Dimensional Magnetism}
\cor{The three-dimensional nature of magnetism is particularly important in a number of areas, which can be identified by their dimensions when compared to the relevant magnetic lengthscales \cite{hubert_book,fruchart_notes}. For example, by patterning curved and three-dimensional {thin} film structures at the nanoscale, with lengthscales on the order of the magnetic exchange length,  control over the  magnetic structure and properties can be obtained via { confinement} of the magnetisation in three-dimensions. On the other hand, in extended - thick, or bulk - systems, which are widely used in technological applications, the internal magnetic structure  is intrinsically three-dimensional: in the absence of lateral confinement that is found in thin film structures, the magnetisation is free to orientate itself in three dimensions, leading to higher degrees of freedom and complex configurations such as three-dimensional spin textures and topological structures.} In this section we present  an overview of the different types of three-dimensional magnetic systems, some of which are discussed in more detail in the following review papers  \cite{streubel16,fernandezpacheco17}.
\subsection{Curved and Three-Dimensional Magnetic Geometries at the Nanoscale}
\cor{In addition to} offering improved physical properties such as high density data storage with the racetrack memory \cite{parkin08}, the introduction of three-dimensionality into a magnetic system, whether it be the introduction of curvature into magnetic thin films \cite{streubel16} or through the patterning of three-dimensional nanostructures \cite{fernandezpacheco17}, can  have a significant influence on the magnetic properties of a system.% \cite{hertel13,hertel16}.

\begin{figure}
    \centering
    \includegraphics[width=\textwidth]{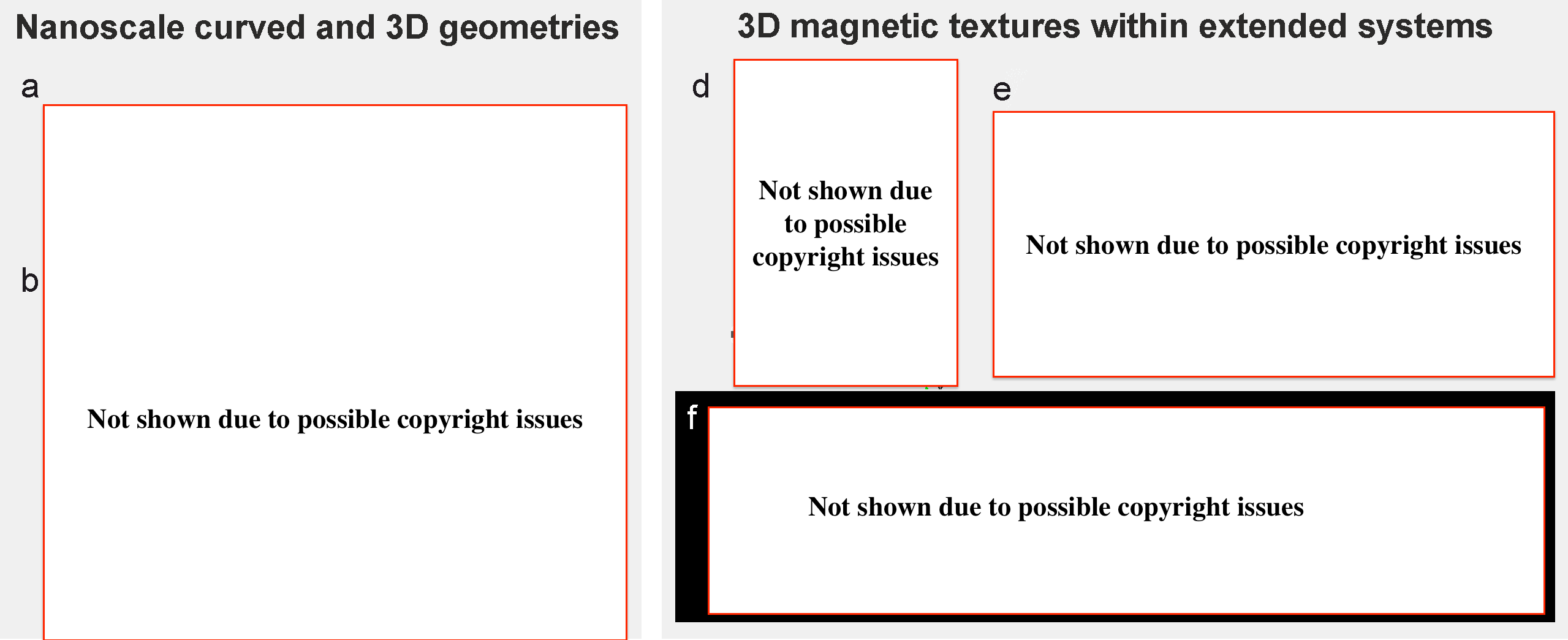}
    \caption{\cor{A brief overview of three-dimensional magnetism, which we consider in two main regimes: nano-patterned curved and three-dimensional magnetic geometries (left) and three-dimensional magnetic textures within extended systems (right). a) Different types of topologically protected domain walls occur in cylindrical magnetic nanowires, which include both the Bloch point domain wall (left) and the transverse domain wall (right) that are depicted schematically. b-c) Micromagnetic simulations of vortex domain walls in hollow magnetic nanotubes (b) predict very high domain wall velocities, leading to magnetochiral effects and reaching the magnonic regime (c) where the spin Cherenkov effect occurs, and spin waves are emitted from the moving domain wall. d) Three-dimensional topological structures such as the surroundings of a circulating Bloch point (pictured) have been observed within extended magnetic materials. Scale bar represents $100\,\rm{nm}$. e) In addition, complex three-dimensional structures formed of bulk magnetic charges have been predicted. Here, isosurfaces are shown that correspond to magnetic charges within the structure, that can be seen to twist in a helix along the axis of the whisker. Another three-dimensional magnetic texture are  hopfions (f), three-dimensional topological solitons that have recently been predicted for magnetic materials. Hopfions are defined by their \textit{Hopf charge}, that quantifies the linking of pre-images corresponding to certain directions of the magnetisation in three dimensions. Here, the blue isosurface corresponds to $\mathbf{m}=[0,0,-1]$ and the red isosurface to $\mathbf{m}=[0,-1/\sqrt{2},-1/\sqrt{2}]$. These curves are seen to intertwine for non-trivial topological hopfions of $Q=3,5$ in the left and right images, respectively.   b) reproduced from Ref.~\cite{yan11}, with the permission of AIP Publishing. c) reproduced from Ref.~\cite{donnelly17}. d) reproduced from Ref.~\cite{arrott16}. e) reproduced with permission from Ref.~\cite{SutcliffePRL17}. Copyright (2017) by the American Physical Society. }}
    \label{fig:hopfions}
\end{figure}

The phenomenon of curved magnetic systems has been of increasing interest in recent years \cite{streubel16,hertel16}, both for their new magnetic properties and due to the prospect of various applications including data storage \cite{parkin08} and shapable magnetoelectronics \cite{makarov16}. By introducing curvature into two dimensional thin films, novel effects such as an effective Dzyalozinskii-Moriya-like interaction  \cite{sheka15,gadidei14,volkov18,streubel16,hertel13} resulting in magnetochirality can be induced. As well as magnetochiral effects, curvature results in an effective anisotropy \cite{sheka15} that acts as an effective magnetic field \cite{gadidei14}, leading to the stabilisation of skyrmion structures \cite{kravchuk16}.  Lattices of three-dimensional curved magnetic shells have been used to achieve a three-dimensional artificial spin ice lattice \cite{May19}.

The natural extension of curved surfaces to three-dimensional structures leads to magnetic nanotubes: elongated closed curved magnetic surfaces that combine the effects of curvature  with the potential for domain wall conduit behaviour \cite{parkin08,allwood05}. A number of analytical works have predicted magnetic phenomena  such as  spin wave non-reciprocity and a {Dzyalozinskii-Moriya}-like dispersion relation for such systems~\cite{otalora16,otalora17}. In addition, magnetic nanotubes are predicted to exhibit rich domain wall dynamics, with magnetochiral ``selection'' in the propagation of vortex domain walls \cite{yan12} and the suppression of the Walker breakdown leading to high domain wall velocities \cite{yan11}, with which new dynamic regimes can be reached. One of these regimes is the magnonic regime, analogous to the supersonic regime for sound, in which the domain wall travels faster than spin waves in the material, leading to the emission of spin waves in a spin-Cherenkov effect \cite{yan11,yan13}. 

\cor{Along with} curvature-based magnetic structures such as magnetic nanotubes, magneto-chiral effects and impressive domain wall properties have also been predicted for magnetic nanowire structures \cite{yan10,stano18}. In these systems, different types of domain wall have been observed, including the Bloch point domain wall, that contains a magnetisation singularity \cite{DaCol14,Wartelle18} and the transverse-vortex domain wall in which the magnetisation points perpendicular to the long axis of the wire \cite{Biziere13,DaCol14,Wartelle18}, shown schematically in Figure~\ref{fig:hopfions}a). When moved using a current or magnetic field, domain walls in cylindrical nanowires are predicted to also exhibit impressive dynamics, with transverse domain walls exhibiting massless properties, thus leading to a complete suppression of the Walker breakdown \cite{yan10}. Contrary to such theoretical predictions, recent experimental investigations have observed the topological transformation of the domain wall structure above a field threshold, highlighting that although very promising, further experimental and theoretical  investigations are necessary to achieve a full understanding of the dynamics of these systems  \cite{wartelle19}.

\subsection{Three-dimensional magnetic textures in extended systems}
In the context of soft magnetic thin films, of thicknesses on the order of the exchange length, where for the most part the spatial anisotropy confines the magnetisation in the plane, a two dimensional model is often sufficient for a description of magnetic behaviour, and a controllable, confined magnetic configuration is often available. As magnetic thin films get thicker, the spatial anisotropy decreases with the result that more complex, higher dimensional magnetic configurations occur. A simple example is that of the magnetic vortex, where the in-plane magnetisation curls around a central core region in which the magnetisation tilts out of plane: in thin films, the core can be approximated as homogeneous throughout the height of the film. \cor{ However, in thicker systems, where the thickness is much greater than the material-specific exchange length ($> 50-100\,\rm{nm}$), where the films are no longer considered ``thin'', the core exhibits a more complex inhomogeneous structure through the thickness of the film, requiring a full three-dimensional treatment \cite{yan07}. }

Indeed, when one moves to larger systems, the complexity of extended systems is reflected in the resulting complicated energy landscapes, which whilst leading to challenges in the understanding and simulation of such systems, can also lead to new nanoscale magnetic phenomena. Within the  complex network of vortices and domain walls  observed in a micrometre-sized ferrimagnetic pillar \cite{donnelly17}, the surroundings of stable singularities of the magnetisation, known as Bloch points, were observed for the first time: { the complex spin texture surrounding one of the observed Bloch points is} shown in Figure~\ref{fig:hopfions}c. In another example, unexpected helical quadrupoles of magnetic charge have recently {been predicted to be a possible source of the  long-observed, yet poorly understood}, domain patterns on the surface of the micrometre-sized Fe whiskers \cite{hubert_book,arrott16}.%, \imp{that as yet are not  well understood}. 

Three-dimensional structures become particularly exciting when one considers their topology. An object with non-trivial topology cannot be smoothly deformed into a trivial, uniform texture, instead requiring a substantial amount of energy to be annihilated, and thus can be thought of as \textit{topologically protected}. A well known example is the skyrmion, a two dimensional planar magnetic object whose spin texture can be mapped onto all directions of a sphere, and which thus has a topological skyrmion number of 1~\cite{braun12}. Topological protection is particularly motivating when it comes to applications such as data storage and logic, where the stability of the textures encoding the data is essential: indeed,  significant research is currently being devoted to the skyrmion racetrack \cite{fert13}, which proposes to exploit both the small size and topological stability of these whirling magnetic structures, \cor{in addition to} the low current-densities needed to propagate them.

Recently there has been a growing interest in  three-dimensional magnetic topological structures, where the corresponding topological number is the \textit{Hopf charge} \cite{Atiyah90,cooper99}, $Q$.
\cor{Also known as  magnetic knots, so-called hopfions are intrinsically three-dimensional topological objects, where their Hopf charge $Q$ quantifies the linking of pre-images corresponding to certain directions of the magnetisation in three dimensions. In particular, in the vicinity of a non-trivial hopfion of \cor{Hopf charge} $Q$, pre-images mapping particular directions of the magnetisation will link with each other $Q$ times, as shown in Figure ~\ref{fig:hopfions}e). In the image shown, the blue isosurface maps regions with the magnetisation direction $\mathbf{m}=[0,0,-1]$ and the red isosurface with $\mathbf{m}=[0,-1/\sqrt{2},-1/\sqrt{2}]$. These pre-images link 3 and 5 times in the left and right images, respectively, corresponding to hopfion charges of $Q=3,5$.}

While at first it appeared challenging to realise  hopfions experimentally, they have been discovered in a variety of materials, including cholesterics~\cite{Bouligand78}, liquid crystals \cite{Chen13,Ackerman2015,Ackerman2017}, anisotropic fluids~\cite{Smalyukh09}, chiral ferrofluids \cite{Ackerman16} and Bose-Einstein condensates~\cite{Hall2016}. Although they have not yet been observed in ferromagnets, there have been multiple theoretical works proposing a number of routes for the realisation of hopfions in a ferromagnet \cite{SutcliffePRL17,Sutcliffe_2018,tai18,liu18,Rybakov19}, and in order to visualise their three-dimensional magnetic configuration, characterisation techniques such as X-ray magnetic tomography will be necessary.

\cor{Besides} fundamental investigations, extended magnetic systems are also important from the point of view of applications. In particular, extended soft magnets are relevant in inductive applications such as motors or sensors that depend on a high permeability \cite{hubert_book,gutfleisch11,streubel15}, while permanent magnets are used for energy harvesting, mechanical and sensor applications \cite{gutfleisch11}. In these materials, the magnetic microstructure is critical to the functionality and thus the performance of the magnet, and it is not only the domain configuration that is relevant, but also the micromagnetic details, which include vortices and magnetic singularities, and the influence of physical defects such as the grain structure of the material. 

As well as improvements in the material properties, increases in device efficiency such as motor output power can be achieved through the design of both the distribution and shape of magnets in three dimensions \cite{Muta04}. Recent advances in the 3D-printing of bonded polymer magnet composites offer the freedom to design permanent magnets in three-dimensions, offering the possibility for optimised field distributions for enhanced efficiency in current applications \cite{Paranthaman2016,huber16,li16}, along with added advantages such as being low weight and low cost, with optimised mechanical properties and corrosion-resistance. Whilst particularly promising for a number of applications,  the magnetic properties of bonded permanent magnets are compromised with respect to sintered permanent magnets \cite{Ormerod97}. A possible solution comes with the recently-developed 3D-printing of permanent magnetic materials such as NdFeB in millimetre-sized complex shapes with micrometer precision~\cite{Jacimovic17}. In this way, significant improvements in the magnetic properties with respect to their bonded counterparts are achieved, with the ability to design bulk systems in three-dimensions, \cor{in addition to} enhanced control over grain size. 

\section{\label{sec:Char3DSt}Characterisation of three-dimensional magnetic nanostructures}
With the growing interest in three-dimensional magnetic systems \cite{streubel16,fernandezpacheco17}, appropriate imaging techniques for the visualisation of the features of three-dimensional magnetic systems are required. \cor{One of the main challenges facing the characterisation of three-dimensional magnetic systems is the \textit{vectorial} nature of the magnetisation. All \textit{three} components of the magnetisation vector field must be determined in three-dimensional space, essentially resulting in a six-dimensional problem. We shall first take a moment to clarify our language: in this review, when referring to three-dimensional magnetisation configurations, the vectorial nature of the magnetisation vector field is assumed. Likewise, \textit{magnetic tomography} refers to the imaging of the three-dimensional vector field in three dimensions.}
 
The three-dimensional magnetisation configuration within the bulk of a magnetic material was investigated for the first time by Libovick\'{y} in 1972, who studied the internal magnetic configuration of a FeSi crystal  indirectly \cite{Libovicky72}. For the specific case of FeSi, heating the sample up to high temperatures results in the formation of platelets that align with the magnetisation of the sample \cite{Libovicky72}. By etching the material in $5\,\mathrm{\mu m}$ steps and imaging the platelet formation, Libovick\'{y} was able to determine the internal magnetic structure of the crystal with micrometre resolution. This technique was extended by Shin \textit{et al.} by combining it with a serial sectioning method involving the etching of the surface in $0.5\,\mathrm{\mu m}$ steps \cite{shin13}, leading to a much higher spatial resolution, with which the microstructure of the domain walls could be determined. Unfortunately, along with being destructive, this technique is specific to FeSi and requires heating the sample to temperatures on the order of $600^\circ\mathrm{C}$, meaning that for other magnetic materials such as soft or permanent magnets with lower Curie temperatures, an alternative method to visualise the internal magnetic microstructure is required.

\cor{Although the first proposals of a non-destructive magnetic tomography technique came in the 1990s \cite{HOCHHOLD1996,badurek97_proposal,badurek97_novel},} it is only in the past decade that magnetic tomography has become a reality. First studies performed with spin polarised neutron tomography were reported in 2008 \cite{kardjilov08}, where the magnetic field within a millimetre-sized superconducting sample was mapped in three dimensions with a spatial resolution of hundreds of micrometres. The magnetisation configuration of a sample, however, was not mapped until 2010, when, with Talbot-Lau neutron tomography with inverted geometry, Manke \textit{et al.}~\cite{manke10} achieved a three-dimensional mapping of the magnetic domain walls, thus determining the size and location of the magnetic domains within the bulk of a FeSi crystal with a spatial resolution of $35\,\mathrm{\mu m}$. Although this implementation of neutron tomography provided the first non-destructive imaging of magnetic domains within a sample, the spatial resolution is limited to lengthscales on the order of tens to hundreds of micrometres. Also, neutron magnetic imaging techniques do not measure the orientation of the magnetisation in a sample, but one rather locates the domain walls~\cite{manke10} or measures the magnetic fields created by the sample magnetization~\cite{kardjilov08}. For three-dimensional nanomagnetic structures and textures that are the focus of this review, a {significantly} higher spatial resolution is required. 

\cor{ The exact spatial resolution required is very material and system dependent. However, one can make the following generalisation: for the study of larger domain configurations, spatial resolutions on the order of hundreds of nanometres are generally sufficient. To gain information on the details of features of the magnetisation configuration, including topological structures such as vortices, skyrmions, and domain walls, a spatial resolution approaching the exchange length of the material (spatial resolution $\approx 5-30\,\rm{nm}$) is required.}

Non-destructive three-dimensional magnetic imaging at the nanoscale has only become a reality in recent years, and was first achieved  with transmission electron microscopy \cite{phatak10,tanigaki15,Wolf2019,wolf2015} and soft X-ray microscopy \cite{streubel15,HierroRodriguez19}.  Although suitable for imaging the magnetic configuration of thin films and nanostructures, the low penetration depth of these techniques means that they are limited to the investigation of samples of total material thickness below approximately $200\,\mathrm{nm}$. In addition, the requirement for vacuum-compatible setups that often lack space, means that obtaining sufficient angular sampling needed for the characterisation of three-dimensional vector fields can be challenging.

A solution for the investigation of thicker extended systems was provided by the use of higher energy X-rays, which offer a higher penetration depth combined with high spatial resolution imaging, as evidenced by recent advances in non-magnetic imaging that have provided nanometre spatial resolution imaging of structures with thickness of the order of many micrometres \cite{holler14,holler17}. Indeed, the first demonstration of hard X-ray magnetic tomography of extended micromagnetic systems allowed for the mapping of the magnetisation vector field within a micrometre-sized GdCo$_2$ pillar with a spatial resolution of $100\,\rm{nm}$ \cite{donnelly17}. Since then, hard X-ray magnetic imaging has also been used to map the domains within systems with strong magnetocrystalline anisotropy \cite{suzuki18}.

For magnetic imaging, however, hard X-rays have certain disadvantages, the main one being   \cor{the limited magnetic contrast, originating from X-ray magnetic circular dichroism (XMCD), that is available}. Indeed, while relatively high XMCD signals are available with soft X-rays, in the hard X-ray regime, however, one indirectly probes the magnetisation, resulting in a significantly weaker magnetic signal \cite{Hippert_book} that limits the spatial resolution, and {poses severe constrains on both the material and the minimal} thickness of a sample that can be measured on a reasonable timescale. {For more details of XMCD, and a comparison of the XMCD signals available in different regimes, please refer to Section \ref{sect:xmcd} and Table \ref{table:xmcd_signal}, respectively.}

When planning measurements for the characterization of three-dimensional vector fields, it is essential to identify the optimal \cor{approach} for {a specific} sample, in order to obtain useful information. We note here that the use of complementary techniques are often useful for a complete understanding of the sample properties and configurations.
In this review article, we will discuss in detail the state-of-the-art available for the study of three-dimensional systems, with a particular focus on X-ray techniques, with a view to providing a clear guide to the current status of three-dimensional magnetic imaging, and how to best make use of it. \cor{Following this discussion, a guide to choosing the optimal experimental parameters for a particular three-dimensional magnetic system is provided in Figure \ref{fig:flow}.}

\section{\label{sec:xrayIm}X-ray magnetic imaging}
In this section we shortly introduce the basis of X-ray magnetic imaging, namely X-ray magnetic circular dichroism, and then go on to describe how it is used to image magnetic domains in magnetic nanostructures.
For a more detailed treatment of the origin of magnetic dichroism the reader is referred to Ref.~\cite{Lovesey96} or Ref.~\cite{Stohr06M}. \\
\subsection{X-ray magnetic circular dichroism}\label{sect:xmcd}
The  dependence of the absorption of light on its polarisation is known as dichroism.
Generally, the response of an electron charge distribution to an electromagnetic wave is 
not isotropic, due, for example, to the molecular orientation or the directionality of chemical bonds.
Depending on the polarization of the incident electromagnetic wave and the local symmetries displayed by the sample, the attenuation coefficient displays linear- or circular dichroism.
Dichroism has also been observed at X-ray wavelengths, though the effect is usually significant only in the vicinity of absorption edge resonances, where the internal fields present into the sample 
strongly affect the excited photo-electrons.
Such photo-electrons are also sensitive  to the magnetisation distribution created by unpaired electrons, leading to the occurrence of X-ray magnetic dichroism. For example, in a magnetic material the absorption might depend on the relative orientation of the X-ray polarization to the preferred magnetic axis. In this case the effect is called X-ray magnetic linear dichroism (XMLD). In the case 
where the absorption depends on relative orientation of the helicity of the X-ray beam and the magnetic easy axis, the effect is named X-ray magnetic circular dichroism (XMCD).

For the study of ferromagnetic materials, the latter effect, XMCD, provides an absorption signal that is proportional to the component of the magnetisation parallel to the direction of propagation of the X-rays. The XMCD signal is dependent also on the particular absorption edge to which the X-ray energy is tuned. 

In the soft X-ray regime resonant X-rays probe electronic transitions between core levels and the magnetically polarised valence band. In particular the ${L_{2,3}}$ edges (2\textit{p}-3\textit{d}) of transition metals and the $M_{4,5}$ edges (3\textit{d}-4\textit{f}) of rare earths directly probe the valence bands 3\textit{d} and 4\textit{f}  of the transition metal and rare earth materials, respectively. This results in relatively high XMCD signals (up to $100\%$)

Whilst in the soft X-ray regime one directly probes the magnetic electrons in the valence level for the $L$ edges of transition metals ($2p-3d$) and the $M$ edges of rare earths ($3d-4f$), in the hard X-ray regime the detection of the magnetic moment is not so straightforward, as it relies on the hybridization between the valence band and the unoccupied orbitals just above it. There are two main types of absorption edge for magnetic materials in this regime: the $L$ edges of the rare earths ($2p-5d$), and the $K$ edges of the transition metals ($1s-4p$), both of which have significantly weaker associated XMCD signals as those found in the soft X-ray regime (see Table \ref{table:xmcd_signal}), especially for 3d transition metals.
Therefore, imaging tools to study magnetic material with X-rays have been developed mostly for the soft X-ray regime, and it is only recently that hard X-ray imaging with nanoscale resolution was achieved \cite{donnelly16}. 
We note here that it has been recently demonstrated that a strong XMCD effect can also be measured  in the hard X-ray regime by performing resonant inelastic X-ray scattering (RIXS), although at the price of a more complex experimental geometry~\cite{SikoraPRL105,SikoraJAP111,InamiPRL119}. 
\begin{table}
\small
	\begin{tabular}{ l  l  l  l  l  l} 
		\hline \hline
		{\normalfont \bfseries X-ray } & \normalfont \bfseries{Absorption edge} & \normalfont \bfseries{\cor{Energy }} & {\normalfont \bfseries  Material } &  \normalfont \bfseries{Example} &  {\normalfont \bfseries Signal strength }  \\ 
		{\normalfont \bfseries type} & {\normalfont \bfseries \cor{(Shells)}}& {\normalfont \bfseries \cor{range}} & {\normalfont \bfseries class}&  {\normalfont \bfseries material}&  {\normalfont \bfseries ($\%$ of $\Delta_{\mathrm{edge}}$)}  \\ 
		\hline
		Soft & $L_{\mathrm{2,3}}$ \cor{(\textit{2p $\rightarrow$3d})} & \cor{$0.4-1\,\rm{keV}$}  & Transition metal & Fe & $100\%$ \cite{harp95}\\ 
		\hline
		Soft &$M_{\mathrm{4,5}}$ \cor{(\textit{2d $\rightarrow$4f})}  & \cor{ $0.9-1.6\,\rm{keV}$}  & Rare earth & Gd & $50\%$ \cite{champion03}\\ 
		\hline
		Hard &$K$  \cor{(\textit{1s $\rightarrow$ 4p})} &  \cor{ $4.5-9.5\,\rm{keV}$ } & Transition metal & Fe & $0.025\%$ \cite{donnelly16}\\ 
		\hline
		Hard &$L_{\mathrm{2,3}}$ \cor{(\textit{2p $\rightarrow$ 5d})} & \cor{$5.7-10.3\,\rm{keV}$} & Rare earth & Gd &  $8\%$ \cite{donnelly16}\\ 
		\hline
		\hline \hline
	\end{tabular}
	\caption{Comparison of the XMCD signals at different absorption edges in the hard X-ray and soft X-ray regimes in terms of $\Delta_{\mathrm{edge}}$, where $\Delta_{\mathrm{edge}}$ is the relative change
in absorption across the absorption edge \cite{donnelly16}. \cor{See Table 1 of Ref. \cite{PAOLASINI2008} for a detailed overview including additional absorption edges.}}
	\label{table:xmcd_signal}
\end{table}

\subsection{X-ray magnetic microscopy in 2D}
When considering imaging the configuration of three-dimensional magnetic structures with X-rays, one first needs to measure a two-dimensional transmission projection of the magnetic structure. To obtain three-dimensional information on the magnetic state, such projections measured at different orientations of the sample can be combined to obtain a three-dimensional image using a  reconstruction algorithm or, for more constrained systems, the projections can be considered analytically, or can be compared with micromagnetic simulations. 

\begin{figure}
    \centering
    \includegraphics[width=\textwidth]{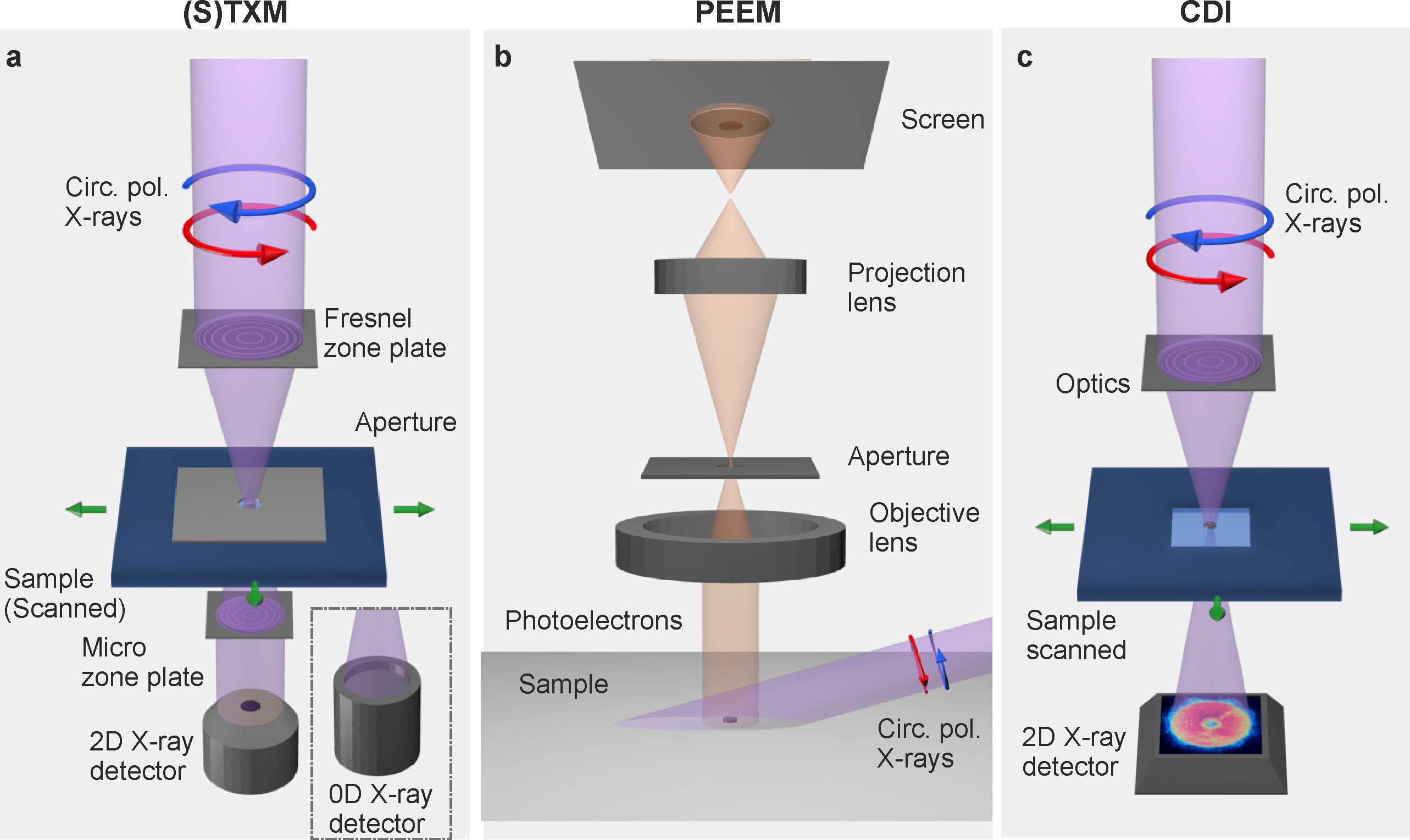}
    \caption{ XMCD based experimental techniques to image magnetic structures with X-rays.\cor{ a) Transmission X-ray microscopy, which can take the form of \textit{full-field} transmission X-ray microscopy (TXM), where the sample is illuminated by a micro-focused beam, and a micro-zone plate is used to produce a magnified image of the sample, which is recorded by a 2D detector. Another variation of transmission X-ray microscopy is \textit{Scanning} transmission X-ray microscopy, where a nano-focused beam is scanned across the sample, and the transmitted X-rays are recorded by a point (0-D) detector (see inset). }b) Photoelectron emission microscopy, and c) Coherent diffractive imaging, where the complex transmission function is reconstructed with an algorithm, with no need for a reference beam. For the case of ptychography, the sample is scanned and diffraction patterns are recorded for different illumination positions on the sample. }
    \label{fig:xrm}
\end{figure}

There are a number of X-ray microscopy techniques that are used to provide a transmission projection, and in this section we briefly illustrate the methods of choice to image magnetic samples.
\begin{enumerate}
    \item STXM \cite{zimmermann18} and TXM \cite{blancoroldan15,streubel15} \\
    \cor{Transmission X-ray microscopy is a photon-in-photon-out technique, in which the intensity of the transmission of a monochromatic X-ray beam through the sample is spatially resolved using X-ray optics. With this technique the energy resolution is given by the beam line monochromator, and the spatial resolution is defined by the X-ray optics.
    There are two main variations of transmission X-ray microscopy: \textit{scanning} transmission X-ray microscopy (STXM) and \textit{full-field} transmission X-ray microscopy (referred to as TXM).}
    
    \cor{In STXM, illustrated in Fig.~\ref{fig:xrm}a, a monochromatic X-ray beam is focused to a small spot size (on the order of tens to hundreds of nanometres) and the X-ray intensity transmitted through the sample is monitored as a function of the position of the  focused beam  on the sample~\cite{Kirz85}. The focused X-ray beam is obtained by using Fresnel zone plates, for which the ultimate achievable spatial resolution is determined by the width of the outermost zones of the zone plate. Spatial resolutions obtained using Fresnel zone plates can typically reach around 25~nm, with demonstrations of spatial resolutions of sub-10~nm achieved recently \cite{rosner18}.}
    
    \cor{For full-field TXM, also shown in Fig.~\ref{fig:xrm}a, analogous to optical microscopes, the focal spot size determines the field of view. A micro-zone plate placed after the sample is then used to produce a magnified sample image, that is  recorded by a X-ray sensitive 2D detector, such as a CCD camera. A spatial resolution of 15~nm has been demonstrated \cite{Chao12}.}
    
    \item X-PEEM and Shadow X-PEEM.\\
    Another X-ray based imaging method is based on the collection of photoelectrons ejected by the sample upon X-ray illumination: photo electron emission microscopy (X-PEEM), illustrated in Fig.~\ref{fig:xrm}b. The sample is illuminated by a monochromatic X-ray beam incident at low angles to the samples surface. For such a microscope the spatial resolution is determined by the electron optics and it is limited by three quantities: spherical aberration, chromatic aberration, and diffraction. In practice, for X-ray excitation of electrons, chromatic aberrations dominate.
    Calculations and experiment show that a spatial resolution of about 20-30~nm can be obtained by X-PEEM. As a surface-sensitive technique, X-PEEM typically probes the top $5-10\,\rm{nm}$ of material in a system, and therefore it is ideally suited to flat samples. In recent years, however, X-PEEM has also been applied to the imaging of three-dimensional magnetic nanostructures, where, by harnessing the low incidence-angle of the X-rays, one images the transmission function of the structure that is projected onto the surface behind it. \textit{Shadow} X-PEEM has been used to successfully image the magnetic structure of a number of three-dimensional magnetic nano- and micro-structures, where the effective increase in spatial resolution due to the  projection of the structure leads to a significant advantage in identifying magnetic features {(see Section~\ref{sec:shadowpeem} for more details and references)}.

    \item Lensless imaging.~\cite{donnelly15,donnelly17}\\
    The previously mentioned techniques, while providing substantially higher spatial resolution than conventional optical techniques, have nevertheless failed to deliver the diffraction limited resolution expected for X-ray wavelengths (e.g. $\lambda \sim 1.8$~nm at the Iron L$_3$ edge). Such limitations, except for the case of X-PEEM, stem from the challenges of producing efficient diffraction-limited X-ray lenses.  Such difficulties have led scientists to  {adapt } coherence-based lensless optical microscopy techniques, such as holography or ptychography, to the X-ray regime, as  illustrated in Figure~\ref{fig:xrm}c. 
    
   Lensless imaging is a set of techniques that record scattering patterns without using lenses and recover the complex field of an object either by interfering the scattering beam with a reference source or via phase retrieval algorithms. 
    Several experiments that exploit the coherence of X-ray synchrotron sources were performed in the last years. \cor{ X-ray holography, where the object transmission is interefered with a reference point source, has been pioneered as the technique allows to obtain an image of the sample in a single shot. Following the first demonstration of X-ray holography of magnetic structures \cite{eisebitt04}, the technique has been successfully applied to time-resolved measurements using both synchrotron \cite{bukin2016,Buettner15} and free electron laser \cite{pfau12_} X-rays, as well as a first demonstration of non-magnetic soft X-ray tomography \cite{guehrs12,guehrs15}.  However, the spatial resolution of X-ray holography is fundamentally limited by the size of the reference source, which is typically on the order of 20-30~nm. }
    
    Significantly higher resolutions have been achieved with coherent diffractive imaging \cor{(CDI)} techniques that do not require a reference. In particular, for ptychography,  multiple diffraction patterns are measured for overlapping illumination spots on the sample, and the complex transmission function is then recovered with a reconstruction algorithm. Ptychography has so far delivered the best spatial resolutions in both the hard X-ray (14~nm in three dimensions \cite{holler17}) and the soft X-ray (5~nm in 2D~\cite{shapiro14}) regimes.
    
    When it comes to magnetic imaging, slightly lower spatial resolutions have been achieved due to the lower intensity of magnetic scattering compared to electronic scattering. \cor{Both coherent diffractive imaging \cite{turner11} and ptychography \cite{tripathi11} have been applied to magnetic imaging, with spatial resolutions of 45~nm \cite{donnelly16} and 12~nm \cite{shi16} with hard and soft X-rays, respectively, having been recently achieved for dichroic ptychography. It is expected that the increase in coherent flux with the next generation of synchrotron light sources will be key to improving the sensitivity and spatial resolutions further.}
\end{enumerate}

\section{\label{sec:3DIm}Three-dimensional X-ray magnetic imaging}
In recent years, with growing interest in three-dimensional magnetic systems, there have been a number of demonstrations of three-dimensional magnetic imaging, and further proposals for experimental and reconstruction techniques. Experimentally there are two main questions when considering X-ray magnetic tomography: that is, the experimental geometry, which determines what part of the vector field is probed during the measurement, and the route to determining the three-dimensional magnetisation vector field. 

In this section we will begin by discussing the different experimental geometries available in Section~\ref{sect:expt_geometries}, and then go on to discuss the use of reconstruction algorithms in determining a three-dimensional vector field in Section~\ref{sect:data_analysis}. Along the way, we will introduce the experimental demonstrations of recent years. 

\cor{At the end of this section, after the parameters for three-dimensional imaging have been discussed, we have included a flow chart, recommending experimental parameters for experimental investigations of three-dimensional magnetic systems, including the experimental geometry, and the X-ray energy suitable for different sample types. This can be found in Figure~\ref{fig:flow}.}

\subsection{Experimental geometries for Three-dimensional magnetic imaging}\label{sect:expt_geometries}
%Having given an overview of the scientific results obtained so far with 3D X-ray magnetic imaging, we now look in more detail at the experimental geometries used, which determine how many components of the magnetisation are probed, and what kind of reconstruction algorithm is needed to retrieve the data. 

There have been four different geometries for three-dimensional magnetic imaging proposed or demonstrated so far, each of which has different advantages and disadvantages in terms of both data quality, and experimental feasibility. \cor{ We note that for the reconstruction of a three-dimensional vector field without \textit{a priori} information, one is limited to geometries in which all three components of the magnetisation are probed, that include the {dual-axis tomography} \cor{(Section \ref{sect:dual_axis}), {effective dual-axis tomography} (Section \ref{sect:eff_dual_axis}) or the laminography (Section \ref{sec:shadowpeem}) geometries. }}

\subsubsection{Tomographic geometry}
The first geometry is the simplest, experimental-feasibility-wise: the standard tomographic geometry \cite{suzuki18,streubel15,blancoroldan15}, which is shown for the case of a free-standing sample in Figure~\ref{fig:geometries}a. In this geometry, the sample can be rotated by $360^\circ$ about a tomographic axis perpendicular to the X-ray beam. When measuring XMCD,  the magnetisation parallel to the direction of propagation of the X-rays is probed, meaning that for a tomographic scan, the two components of the magnetisation in the plane perpendicular to the rotation axis are measured, and the measurement is \textit{not} sensitive to the third component parallel to the rotation axis. 

This geometry has been used successfully for cases in which the magnetisation is well defined along a particular direction. Streubel \textit{et al.} used soft X-ray tomographic imaging to determine the influence of curvature on magnetic thin films by imaging the three-dimensional structure of magnetic microtubes, which consisted of magnetic thin films of varying anisotropies being rolled up, as shown in Figure \ref{fig:3Dmagimaging}a. To determine the magnetic structure, an algorithm was designed for hollow cylindrical samples of known magnetic anisotropy that considered the difference in contrast between neighbouring angular projections. In this way, the domain structure on the surface of the cylinder could be determined for samples consisting of one and more windings, allowing for further insight to the role of magnetostatic coupling in the  formation of the magnetic domains \cite{streubel15}.

We note that for the case of a free-standing sample such as the cylinder shown schematically \cor{in Figure~\ref{fig:geometries}a}, this geometry works well. However, for a sample mounted on a SiN$_4$ membrane (Figure~\ref{fig:geometries}b), the frame of the membrane will result in shadowing of the X-ray beam at a number of angles,  thus {causing} a so-called ``missing wedge'' {in the collected data} \cite{Hierro-Rodriguez18}. This was the case in the determination of the structure of a $\rm{NdCo_5}$ film with weak perpendicular anisotropy capped with layers of permalloy by Blanco-Roldan \textit{et al.} \cite{blancoroldan15}, where the thin film was mounted on an X-ray-transparent membrane (Figure \ref{fig:3Dmagimaging}b). By measuring XMCD projections about a tomographic axis, as shown in { Figure \ref{fig:geometries}a} and fitting the angular dependence of the XMCD signal, they were able to determine the presence of multiple in-plane and out-of-plane domains superimposed on one another, that occur {due to the canting of the magnetisation}. Closer consideration of the details of the domain structure led to the identification of topological defects such as merons - i.e. a ``half-skyrmion'' - at the dislocations of the stripe domain pattern \cite{blancoroldan15}. \cor{We note that although the tomographic geometry was used to determine the vectorial nature of the magnetisation, the magnetic structure was not mapped in three dimensions, and so this is not an example of magnetic tomography, as defined in Section~\ref{sec:Char3DSt}.}

In addition to soft X-rays, the standard single-axis tomographic geometry has also been demonstrated with hard X-rays where, by using an adapted form of a conventional tomographic reconstruction algorithm, Suzuki \textit{et al.} were able to determine the three-dimensional magnetic internal structure of a micrometre-sized GdFeCo disc with uniaxial magnetic anisotropy \cite{suzuki18}. More details of the results, and the reconstruction algorithm used, are given in \cor{ Section~\ref{sect:tomo_recons}.}

\subsubsection{Dual-axis tomography}\label{sect:dual_axis}
A second geometry is the dual-axis tomographic setup shown in Figure~\ref{fig:geometries}c,d and proposed by \cite{Hierro-Rodriguez18} \textit{et al.}, where magnetic tomography was demonstrated with numerical simulations. In this case, projections are measured around two tomographic axes at $90^\circ$ to each other. As two components will be measured for each rotation axis, all three components of the magnetisation are probed in a dataset, meaning that a tomographic reconstruction of the three-dimensional magnetisation vector field is possible, as has been demonstrated by Hierro-Rodriguez et al \cite{Hierro-Rodriguez18,HierroRodriguez19} and which will be discussed in more detail in~\cor{ Section~\ref{sect:tomo_recons}}. For the case of a free-standing sample, rotating the sample $360^\circ$ around both axes could be challenging due to either the sample holder blocking the beam at some angles, or due to an asymmetric sample such as the cylinder shown in one direction absorbing too much for some rotations. As with the previous geometery, when used with a sample mounted on a membrane, there will be a missing wedge in the measured angles, which will have adverse effects on the reconstruction \cite{Hierro-Rodriguez18}.

\subsubsection{Effective dual-axis tomography}\label{sect:eff_dual_axis}
A third demonstrated geometry is an effective dual-axis tomographic setup, demonstrated by Donnelly \textit{et al.} in~\cite{donnelly17,donnelly18} and shown in Figure~\ref{fig:geometries}e, in which the sample is orientated at $0^\circ$ and $30^\circ$ to the rotation axis, and all three components of the magnetisation are probed with the two tomographic measurements. This geometry has the advantage that the issues with the previous dual-axis setup associated with the high absorption of an asymmetric sample, and shadowing of the beams, are easier to avoid with a standard tomographic holder and sample. This geometry was used in the first demonstration of X-ray magnetic tomography~\cite{donnelly17} where, using a tailored reconstruction algorithm \cite{donnelly18}, the three-dimensional internal magnetic configuration of a micrometre-sized magnetic pillar was resolved. The reconstruction algorithm, and the experimental results of this work will be described in the following section.

\subsubsection{\label{sec:shadowpeem}\cor{3D} Shadow X-PEEM/ Laminography}
\begin{figure}
    \centering
    \includegraphics[width=\textwidth]{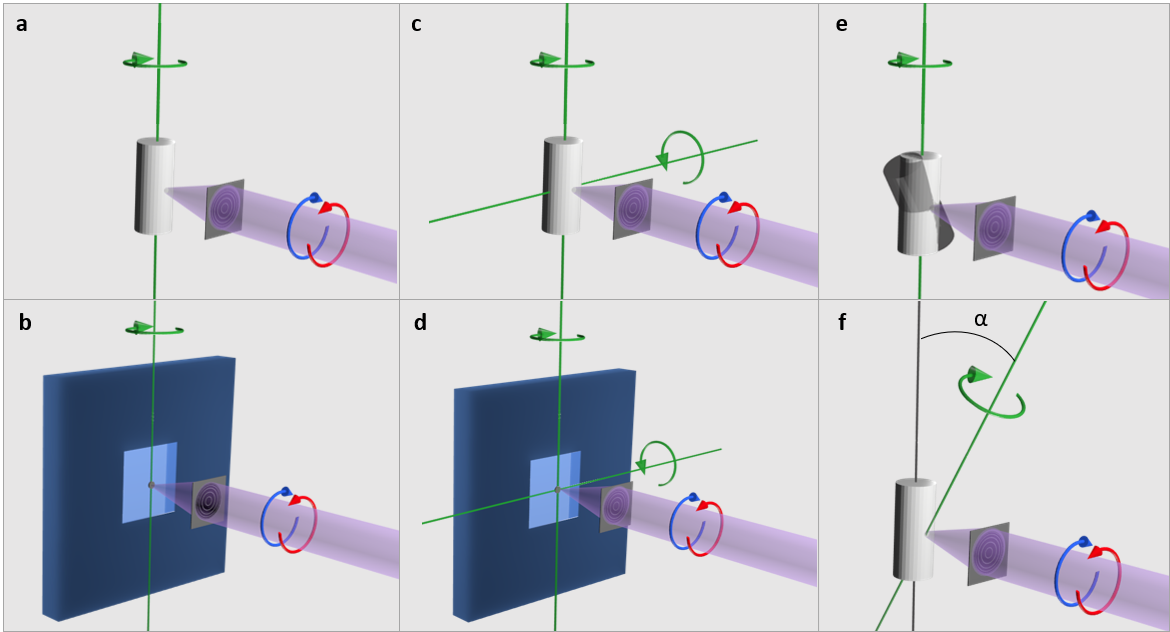}
    \caption{Experimental geometries for three-dimensional magnetic imaging. a,b) a standard tomographic geometry, shown for the case of a free-standing object (a) (used in \cite{streubel15}) for which projections can be measured about $360^\circ$, and for a SiN$_4$ membrane (b) (used in \cite{suzuki18}), for which case the frame of the membrane restricts the angles over which projections can be measured, leading to a \textit{missing wedge}. In the single axis geometry, only two components of the magnetisation are measured. c,d) a dual axis tomographic geometry for a free-standing object (c) and a membrane (d) (proposed in \cite{Hierro-Rodriguez18}). The dual-axis geometry results in the measurement of all three components of the magnetisation, and again the use of a membrane results in a missing wedge in the acquired data. e) an effective dual-axis tomographic setup, where a tomogram is measured for the sample in two different orientations, at $0^\circ$ and $30^\circ$, (used in \cite{donnelly17}). In this setup, all three components of the magnetisation are measured with two tomographic measurements. f) a laminography geometry, where the rotation axis is rotated at an angle $\alpha$ with respect to the beam ($\alpha=90^\circ$ is the tomography geometry). For $0<\alpha< 90^\circ$, all three components of the magnetisation are measured with \textit{one} rotation axis.  }
    \label{fig:geometries}
\end{figure}

\begin{figure}
    \centering
    \includegraphics[width=\textwidth]{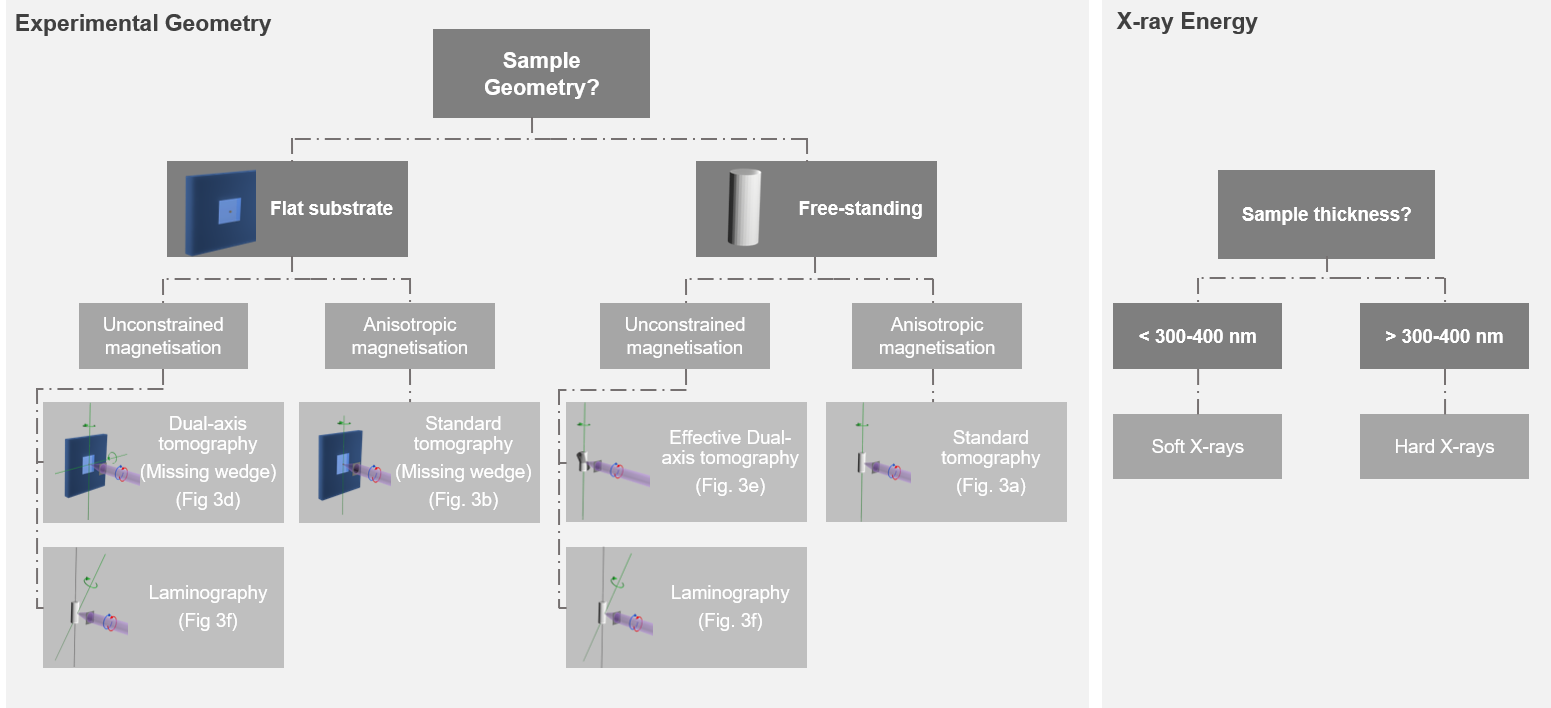}
    \caption{\cor{ Flow chart indicating the recommended experimental parameters for the investigation of a three-dimensional magnetic system with X-rays. In particular the experimental geometry (left) and the X-ray energy (right) are considered. An anisotropic magnetisation refers to one that exhibits a known preferred direction, meaning that fewer rotation axes are needed, whilst an unconstrained magnetisation is free to  point in all directions. Experimental geometries shown are discussed in more detail in Section~\ref{sect:expt_geometries} and shown more clearly in Figure~\ref{fig:geometries}.  }}
    \label{fig:flow}
\end{figure}

An alternative geometry  in which the rotation axis is not perpendicular to the direction of propagation of the X-rays, but instead is at an angle ($90-\alpha $) to the X-ray direction, as shown in Figure \ref{fig:geometries}f, is implemented in   \textit{3D Shadow} X-PEEM. \cor{For X-PEEM, this angle $\alpha$ is fixed to a certain value depending on the PEEM microscope design (for the Swiss Light Source, $\alpha=16^\circ$).} We note here that, even with one projection, shadow X-PEEM has been useful in determing the magnetic state of magnetic nanowires~\cite{kimling11,DaCol14,bran17,kan2018,wartelle19} and magnetic nanotubes~\cite{stano18,jamet15,wyss17} due to the relatively simple configurations that occur in these confined geometries, and due to the possibility to extract extra information: by imaging the structure directly, and in transmission, it is possible to obtain a combination of surface and volume information. In this way, not only has the magnetic domain structure been determined, but the \textit{type} of domain wall -  Bloch point or transverse-vortex  - has been identified~\cite{DaCol14}. \cor{For the determination of three-dimensional vector fields, however, a three-dimensional imaging technique consisting of measuring multiple projections at different sample orientations - i.e. 3D Shadow X-PEEM - is required.}

\cor{3D Shadow X-PEEM has been achieved experimentally by rotating the sample and measuring multiple projections to obtain three-dimensional information by Streubel \textit{et al.} for the study of magnetic microtubes \cite{streubel15}.} Again, similar to the study of microtubes in the ``standard'' tomographic geometry, an algorithm that assumed hollow cylindrical samples of a particular magnetic anisotropy was used, that considered the difference in contrast between neighbouring angular projections, and allowed for the recovery of the magnetic domain distribution on the surface of the tube.

 \cor{A similar geometry with $0<\alpha<90^\circ$, that is combined with transmission imaging, is known as the \textit{laminography} geometry. \cor{ As well as providing high spatial resolution imaging of extended systems \cite{Holler19}, as will be discussed in Section~\ref{sect:future}, for certain $\alpha$, in the laminography geometry all three components} of the magnetisation are probed with \textit{one} axis of rotation, therefore  providing the possibility to circumvent the problems  of the missing wedge  associated with standard tomographic geometries. A first demonstration of magnetic laminography, and its extension to time-resolved studies of three-dimensional magnetisation dynamics, highlights the suitability for laminography as a flexible alternative to more standard magnetic tomography geometries \cite{donnelly19}.}

%\subsection{Combining multiple }
%We can divide the experimental demonstrations of the 3D X-ray magnetic imaging in two parts according to the scale of three-dimensional system that was investigated, and the {photon energy} of X-rays used. 
%\subsubsection{Soft X-rays and thin-film systems}
%First demonstrations of 3D X-ray magnetic imaging were performed with soft X-rays, where magnetic thin film systems were investigated by measuring projections at a number of different angles with respect to the X-ray beam  \cite{streubel15,blancoroldan15}. Although a tomographic reconstruction algorithm was not implemented, three-dimensional information on the magnetic state was obtained by fitting the angular dependence of the magnetic signal, and therefore extracting the orientation of the magnetic vector field.

\begin{figure}
    \centering
    \includegraphics[width=\textwidth]{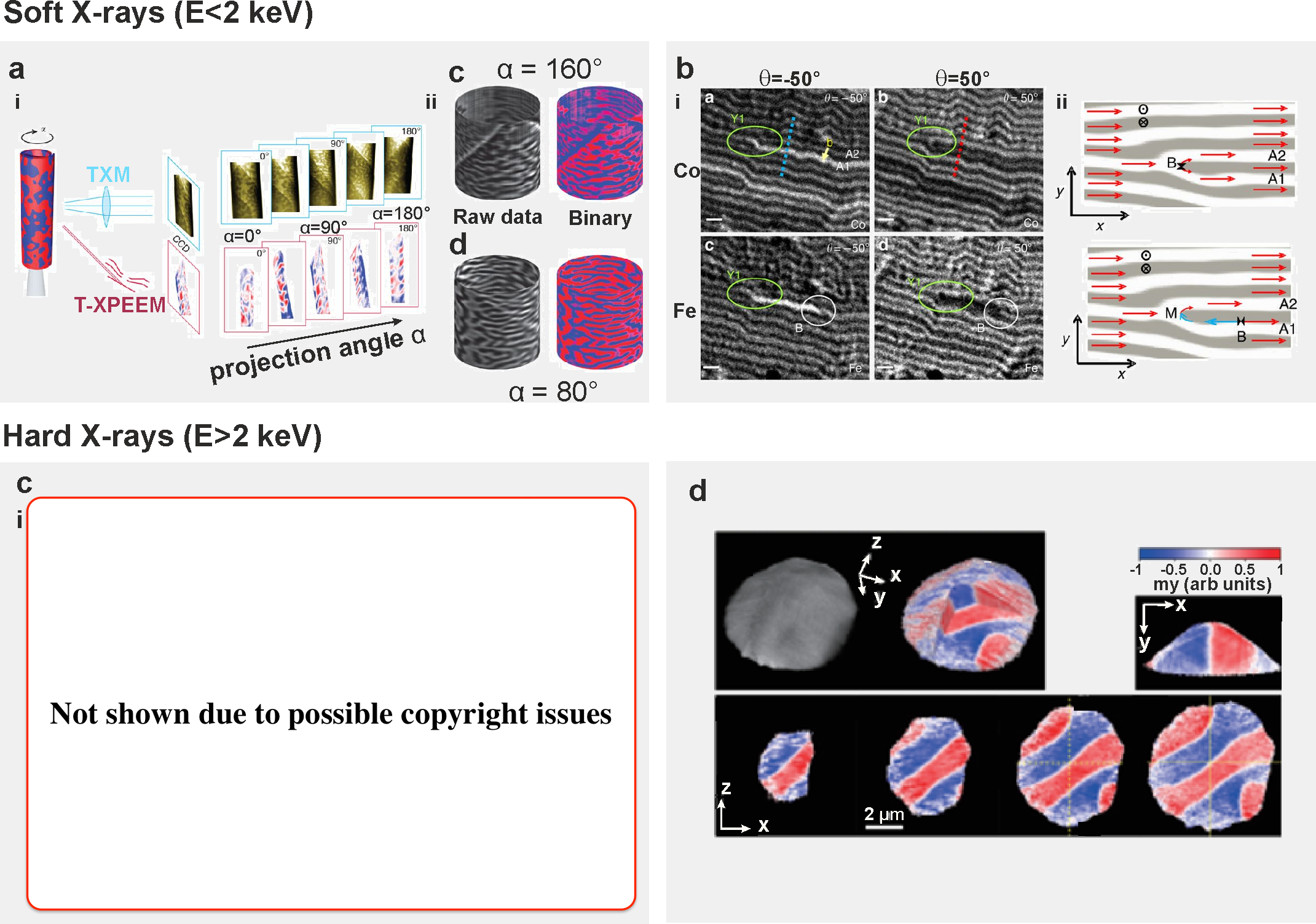}
    \caption{ \cor{Experimental demonstrations of the imaging of magnetisation vector fields with a,b soft X-rays and c,d hard X-rays. a) (i) The magnetic configuration of a magnetic microtube is determined in three dimensions by measuring 2D projections of the magnetic structure at different orientations of the sample with respect to the X-ray beam. (ii) Using an adapted algorithm, the distribution of the domains on the three-dimensional tube was determined. b) Details of the  domain structure of a planar magnetic thin film with canted magnetisation are determined by analysing the angular dependence of the magnetic contrast. (i) By comparing XMCD Images of the magnetic structure at the Co  and Fe edges at different angles, (ii) a dislocation core can be identified, at which the presence of a Bloch point (B) and a Bloch-point - Meron pair (M-B) is determined. c) The internal magnetic configuration of a micrometre-sized GdCo pillar is determined. (i)  The three-dimensional magnetic vector field is reconstructed using a tomographic algorithm, revealing a network of topological defects such as vortices (blue dots) and antivortices (red dot) and (ii) Bloch points . d) The internal structure of a GdFeCo disc with strong uniaxial anisotropy is determined using a single-axis tomographic setup. With an adapted tomographic reconstruction algorithm, a single component of the magnetisation is reconstructed, revealing stripe domains through the depth of the disc. Images in a) are reproduced from \cite{streubel15}, b) from \cite{blancoroldan15}, c) from \cite{donnelly17} and d) from \cite{suzuki18}.}}
    \label{fig:3Dmagimaging}
\end{figure}

\subsection{Recovering the vector field: data analysis}\label{sect:data_analysis}

One of the main challenges for magnetic tomography has been the reconstruction of three-dimensional magnetic vector fields with an appropriate reconstruction algorithm. In normal computed tomography, a single value is reconstructed for each voxel within the structure, resulting in a well-posed problem. For magnetic tomography on the other hand, not one, but three components of the magnetisation have to be recovered for each voxel, which results in requirements both for the type of dataset that is measured - i.e. that all three components of the magnetisation are probed - as well as challenges for the reconstruction. 

Indeed, XMCD projections probe the component of the magnetisation parallel to the X-ray direction, and thus represent a \textit{longitudinal probe} \cite{braun91}. For a tomographic dataset measured about a single axis of rotation (shown schematically in Figure \ref{fig:geometries}a), only the in-plane components of the vector field are measured - meaning that in order to be obtain all three components of the magnetisation, additional information in the form of data measured around additional tomographic axes or prior information is required. 

As discussed in more detail in \cite{donnelly18,donnelly_thesis}, being sensitive to the components of a vector field does not guarantee that they can be correctly reconstructed. Indeed, one of the limitations of a longitudinal probe is that divergent structures do not give a net signal, and thus prove difficult to measure. For a magnetic tomographic dataset about a single rotation axis, it has been shown both mathematically \cite{braun91} and with numerical simulations \cite{donnelly_thesis,donnelly18} that only the rotational part of the vector field in the plane perpendicular to the rotation axis can be recovered. For the recovery of the divergence of the magnetisation in the plane, additional constraints such as on the magnitude of the magnetisation \cite{donnelly_thesis} or more measurements about additional axes of rotation can be applied \cite{donnelly18}. Higher errors in the reconstruction have been observed in the vicinity of highly divergent structures such as Bloch points and vortex cores \cite{donnelly18}.

\subsubsection{Tomographic reconstruction algorithms for a magnetisation vector field}\label{sect:tomo_recons}
The first demonstration of a full tomographic reconstruction of the three-dimensional vector field was performed by Donnelly \textit{et al.} using hard X-rays, where the internal three-dimensional magnetic structure of a magnetic micropillar was mapped out, revealing a network of topological structures including vortices, domain walls and Bloch points \cite{donnelly17} (Figure \ref{fig:3Dmagimaging}c). In that work,  the effective dual-axis experimental geometry detailed in Figure \ref{fig:geometries}e was used. A ``2-step'' gradient-based iterative reconstruction algorithm was used to reconstruct the two components of the magnetisation in the two planes perpendicular to the rotation axis, and the three-dimensional magnetic structure was then obtained by iteratively solving a set of simultaneous equations.

This reconstruction method was upgraded to a more versatile ``arbitrary projection'' approach in \cite{donnelly18}, where all tomographic projections are combined to recover the three components of the magnetisation in a single iterative reconstruction. \cor{In each iteration, a scalar error metric is calculated by calculating the difference between the measured projections, and the projections of the current reconstructed object. An analytical expression for the gradient of the error metric with respect to each variable (i.e. the non-magnetic, and magnetic components) is used to direct the update of each variable. These spatially resolved gradients are calculated during  each iteration, thus allowing for a more direct convergence of the reconstruction.}   It was shown that by combining all projections together in a single step, this new technique provides a more accurate reconstruction of the magnetisation with lower errors \cite{donnelly18}, and is not limited to tomographic geometries. Using numerical simulations of magnetic tomography of a complex micromagnetic simulated structure, the technique was demonstrated to be robust for a variety of magnetic configurations. In particular, the reconstruction of divergent configurations such as Bloch points was demonstrated, showing that although divergences of the magnetisation prove challenging, a good reconstruction can still be obtained \cite{donnelly18}. The reconstruction algorithm is {freely} available at \cite{claire_donnelly_2018_1324335}.

A second example of hard X-ray magnetic tomography was performed by Suzuki \textit{et al.} for the case of a magnetic material with uniaxial anisotropy. By assuming that the magnetisation was aligned along a single direction, they were able to adapt the filter-back projection algorithm that is commonly used for scalar tomography, and thus reconstruct the single component of the magnetisation along the anisotropy axis. In this way, the magnetic stripe  domains within a large GdFeCo disc were imaged with diameter $10\,\rm{\mu m}$ and thickness $5\,\rm{\mu m}$ with perpendicular anisotropy~\cite{suzuki18}, as shown in Figure \ref{fig:3Dmagimaging}d. \cor{As hard X-rays were used, the authors were able to reveal that the stripe domains remain straight through the bulk of the material, even on a micrometre length scale.}

\cor{Recently Hierro-Rodriguez \textit{et al.} have proposed an alternative iterative tomographic reconstruction, which is based on solving a set of linear equations using an algebraic technique \cite{Hierro-Rodriguez18}. In particular, for each pixel on the detector, a linear equation is formulated which expresses the measured intensity as a function of the three-dimensional object (including e.g. the non-magnetic, and magnetic components). A sparse matrix combining the equations for all pixels on the detector is calculated for each projection angle, and the three-dimensional object reconstructed iteratively using the algebraic reconstruction technique (ART)\cite{Kak_tomo_book}.} The authors propose this reconstruction method specifically for soft X-ray magnetic tomography, and demonstrate the reconstruction of magnetic structures thinner than $300\,\rm{nm}$. Interestingly, they determine that the three-dimensional magnetic structure is still well reconstructed even with a missing wedge in tomographic measurements, although there are associated increases in the error of the reconstruction as well, \cor{with the reconstructed vector components being $5-20\%$ lower than the original structure. }This result is very encouraging for soft X-ray magnetic tomography, as the missing wedge (see Figure~\ref{fig:geometries}b) is difficult to avoid experimentally for samples mounted on standard sample holders such as silicon nitride membranes and {Omniprobe} {transmission electron microscope} holders in this geometry. Recently the same authors have published an experimental demonstration of soft X-ray magnetic tomography using the proposed dual-axis geometry where the three-dimensional magnetic structure of a ferromagnetic film was determined, revealing the three-dimensional structure of  closure domains, and the presence of meron-like singularities~\cite{HierroRodriguez19}.

\cor{We note that for the reconstruction algorithms used by Donnelly \textit{et al.} in \cite{donnelly17,donnelly18} and by Hierro-Rodriguez \textit{et al.} in \cite{Hierro-Rodriguez18,HierroRodriguez19}, no \textit{a priori} information about the sample or its magnetic properties is required, whilst the demonstration by Suzuki et al. \cite{suzuki18} assume a uniaxial magnetic anisotropy. }

\section{\label{sec:FutPr}Future prospectives}\label{sect:future}
With such progress in the imaging of three-dimensional magnetic systems with X-rays in recent years, both experimentally \cite{streubel15, donnelly17, suzuki18,HierroRodriguez19} and  with the development of new, robust reconstruction algorithms \cite{Hierro-Rodriguez18,donnelly18}, in the near future we expect  X-ray magnetic tomography to become a user-friendly technique available at multiple synchrotron light sources around the world for the imaging of a large variety of magnetic systems.

As this is a new and growing field, there are a number of opportunities for significant technical progress, which will make it possible to address a number of pressing scientific questions, both of which we will discuss briefly in this section.

\subsection{Technical prospectives}
With upcoming advances in X-ray optics and the advent of upgraded and fourth generation synchrotrons, the coherent flux available in the hard X-ray regime can be expected to increase by 2-3 orders of magnitude in the near future \cite{eriksson14,thibault14}. This increase in coherent flux will have direct implications for coherent diffractive imaging techniques such as dichroic ptychography, where it is the available coherent flux that limits the achievable spatial resolution, and can be expected to bring hard X-ray magnetic tomography down to a spatial resolution of the order of $20\,\mathrm{nm}$. 
This increase in coherent flux can be used not only to increase the spatial resolution, but also to decrease measurement times. The magnetic tomography experiment presented in \cite{donnelly17} took around $52\,\mathrm{hours}$; an increase in flux could reduce this lengthy experiment to significantly shorter, more reasonable timescales, opening the way to performing systematic studies involving, for example, field and temperature dependent behaviour, within the time constraints of a synchrotron beamtime of a few days. 

\cor{In addition to lensless imaging techniques such as ptychography, which have been pioneered for magnetic tomography up until now, full-field and scanning transmission X-ray microscopy represents a promising route for the high spatial resolution imaging of three-dimensional magnetisation vector fields due to the high levels of non-coherent flux available. Although the first demonstration of hard X-ray magnetic tomography with STXM had a spatial resolution in three dimensions of approximately $300\,\rm{nm}$ using a beam diameter of approximately $150\,\rm{nm}$, the use of improved X-ray optics could lead to spatial resolutions below $50\,\rm{nm}$ for both full-field \cite{andrews11} and scanning \cite{Yamauchi_2011} X-ray transmission microscopy. }

 Long term, as tomographic measurements become faster, four dimensional magnetic imaging with nanoscale resolution will become realistic. For example, a time resolved tomographic measurement could be achieved using a pump-probe measurement scheme \cite{beaurepaire96,stoll04,pizzini03,Buettner15,bukin2016,pfau12_}  in which the sample is excited using current, magnetic field or laser pulses. Pump-probe stroboscopic measurements with 2D magnetic X-ray imaging are routinely used to investigate vortex dynamics \cite{raabe05,wohlhuter15,bukin2016}, domain wall dynamics \cite{finizio19}, and the dynamics of Bloch points \cite{Im19} as well as spin waves \cite{wintz16}. In the same way, tomographic projections of the sample could be measured at a number of delay times with respect to the exciting pulse, and thus time-resolved tomographic reconstructions obtained. In this way, complex three-dimensional magnetisation structures and the role of their spatial evolution and interactions in relaxation processes could be explored, along with, for example, their interaction with physical defects. \cor{To sufficiently explore a parameter space within a beamtime, reasonable measurement times for a single three-dimensional image are required. For example, for a typical beamtime of 15 shifts\footnote{One shift corresponds typically to 8 hours of measurement time.}, and measuring 8-10 timesteps, an upper limit for the scan time would be on the order of one beamtime shift per three-dimensional image. Such scan times will require significant advances in both instrument capabilities, such as the available coherent flux, and  the development of a pump-probe  tomography setup.}
 
 The tomographic reconstruction techniques presented in \cite{donnelly17,donnelly18,Hierro-Rodriguez18} are not limited to hard X-ray measurements for which the magnetic signal is very weak. For suitably thin systems, one could directly combine the reconstruction algorithms described in \cite{donnelly18,Hierro-Rodriguez18} with high spatial resolution soft X-ray imaging that is provided by current transmission X-ray microscopy and holographic techniques. In particular, with single digit nanometre spatial resolution soft X-ray magnetic imaging already having been reported with STXM \cite{rosner18}, this combination of high spatial resolution soft X-ray imaging with the iterative magnetic tomography reconstruction algorithm offers a route to the determination of magnetic structures on length scales below $10\,\mathrm{nm}$. 
 
 For the study of flat systems and magnetic thin films, the proposed  tomographic setup that have been demonstrated experimentally result in a ``missing wedge'', meaning that for certain tomographic angles no projections would be measured, and which leads to  a loss of accuracy in the resulting reconstruction. 
 To avoid this problem, alternative geometries can be considered. In particular, in laminography, where the rotation axis is not perpendicular to the probing X-ray beam, access to all three components of the magnetisation vector field is obtained with a single axis of rotation, and there is no resulting missing wedge. As laminography is ideal for the study of flat and thin samples, it is directly compatible with high spatial resolution soft X-ray imaging, and with lithographically patterned samples on a membrane, providing a flexible setup for in-situ and pump-probe studies \cite{donnelly19}.

\subsection{Scientific}

The discussed future technical advances can be expected to provide i) spatial resolutions approaching the exchange length, ii) significantly faster measurement times, and iii) the possibility for \textit{in-situ} experiments. 
These future capabilities will open the door to answering a number of scientific questions, a few of which we will summarise here.

Within curved and three-dimensional patterned magnetic nanostructures, there are a number of open questions which three-dimensional magnetic imaging will be useful to help answer. The majority of investigations of three-dimensional systems has so far concerned single, isolated magnetic nanowires, and magnetic micro- and nanotubes. As fabrication capabilities for more complex three-dimensional structures improve \cite{donnelly15,May19,Pablo-Navarro17,Winkler19,Sanz-Hernandez18_ACSNano,keller18}, the next steps towards exploiting the new effects such as curvature-induced effects %and shape-induced frustration 
will involve moving to more complex geometries and coupled systems. In more complicated systems, where new spin textures and magnetic phenomena are predicted to occur, understanding the details of the magnetic configuration will require three-dimensional imaging techniques as discussed in this review.
In addition, one of the most promising avenues for patterned magnetic nanostructures is in their rich dynamics. Future developments of time-resolved three-dimensional magnetic imaging, whether it be pump-probe or quasi-static, will be key to understanding the complex domain wall dynamics, both those that are predicted by theoretical studies \cite{yan10,yan11,yan12,hertel16}, as well as unexpected effects such have been recently discovered experimentally \cite{wartelle19}.
    
The fundamental understanding of complex magnetic textures, such as skyrmion lattices and hopfions, will also benefit significantly from new capabilities in three-dimensional magnetic imaging. Skyrmion lattices have been observed in several materials by means of scattering techniques \cite{Zhang2018} or electron microscopy \cite{Tonomura2012,park2014,Rajeswari2015}. However, scattering techniques offer only an average picture of the magnetic moments configuration in the material, \cor{while transmission electron microscopy is notoriously limited to thin films, with the result that both sample preparation, and beam damage during imaging, are challenging. As a result,} obtaining a detailed understanding of the magnetic structure within the bulk of a lattice has until now not been possible. To gain a quantitative description of the role of defects both in the formation, and the dynamical behaviour, of skyrmion lattices and the phase transitions between skyrmionic phases~\cite{Fujishiro2019}, a technique offering high spatial resolution imaging of the three-dimensional magnetic configuration is essential. 
Such techniques {will be} even more relevant for the observation of higher dimensional topological magnetisation configurations such as the hopfions, which are not predicted to occur in a periodic arrangement, and will have an intrinsically three-dimensional magnetic structure.

As  X-ray magnetic tomography approaches single digit nanometre spatial resolution, on the order of the exchange length of many magnetic materials, a number of routes for the investigation of fundamental magnetic systems will become possible, such as the observation of magnetic textures in crystals with complex magnetic phase diagrams that are characterized by incommensurate magnetic modulation and/or the presence of strong magnetic diffuse scattering, such as TmB$_4$~\cite{Siemensmeyer08}.

Higher spatial resolutions will also enable the \cor{non-destructive} study of permanent magnets. Researchers in this field have so far struggled to characterize their sample due to the lack of a truly three-dimensional magnetic characterization methods. Such materials are often characterised by thinning the samples to thicknesses on the order of tens to hundreds of nanometres \cite{ono11}, which, although giving insight to the magnetic and material properties of the sample, results in significant changes in the magnetic configuration. By using hard X-ray three-dimensional magnetic imaging, it will be possible to characterise samples of sizes on the order of tens of micrometres, thus greatly reducing the requirements for sectioning, and limiting the modification of the magnetic state.

\section{Conclusions}
In this text we have reviewed the state-of-the-art of magnetic nanotomography imaging for three-dimensional magnetic systems. This young yet very promising technique will  benefit tremendously from the advent of diffraction-limited X-ray sources, and will provide quantitative mappings, with single digit nanometre spatial resolution, of three-dimensional magnetic moment configurations in a wide variety of material systems, ranging from three-dimensional nanostructures to bulk materials. At the nanoscale, three-dimensional magnetic nanostructures promise to deliver new physics, which will lead to breakthroughs in the next generation of devices for sensing, and manipulating and storing information. Within larger-scale systems, bulk crystals are  of great scientific interest both for materials that are predicted to host exotic topological configurations, as well as inductive and permanent magnets that are relevant for green energy applications. 

We expect that  these new capabilities for the characterisation of three-dimensional magnetic systems  will lead to many exciting  discoveries in next years, that will be relevant both for fundamental physics, but also to the development and optimisation of sustainable technological applications  in our day-to-day lives.  
\section*{References}

\bibliographystyle{ieeetr}

\end{document}